# Analysis of Application Delivery Platform for Software Defined Infrastructures

**Lav Gupta* and Raj Jain**

Department of Computer Science and Engineering,
Washington University in St. Louis,
St. Louis, MO 63130
jain@wustl.edu
*Corresponding author

**Mohammed Samaka**

Department of Computer Science and Engineering,
Qatar University,
Doha, Qatar.
samaka.m@qu.edu.qa

***Abstract—*** Application Service Providers (ASPs) obtaining resources from multiple clouds have to contend with different management and control platforms employed by the cloud service providers (CSPs) and network service providers (NSP). Distributing applications on multiple clouds has a number of benefits, but absence of a common multi-cloud management platform that would allow ASPs dynamic and real time control over resources across multiple clouds and interconnecting networks makes this task arduous. Open Application Delivery Network (OpenADN), a multi-cloud management and control platform, fills this gap. However, performance issues of such a complex, distributed and multi-threaded platform, not tackled appropriately, may neutralize some of the gains accruable to the ASPs. In this paper, we establish the need for and methods of collecting precise and fine-grained behavioral data of OpenADN like platforms that can be used to optimize their behavior to control operational cost, performance (e.g., latency) and energy consumption.

Lav Gupta is a senior member of IEEE. He received a BS degree from Indian Institute of Technology, Roorkee, India in 1978 and MS degree from Indian Institute of Technology, Kanpur, India in 1980. He is currently pursuing Ph.D. degree in Computer Science & Engineering at Washington University in St Louis, Missouri, USA. He has worked for about 15 years in the area of telecommunications planning, deployment and regulation. With the sector regulatory authority, he worked on technology and regulation of next generation networks. He has also worked as a senior teaching faculty of Computer Science and Access Network Planning for a number of years in telecommunications academies in India and the UAE. He is the author of one book, five articles and has been a speaker at many international seminars. He was the recipient of the Software Award from Computer Society of India in 1982 and best faculty award at Etisalat Academy, UAE in 1998.

Raj Jain is a Fellow of IEEE, a Fellow of ACM, a Fellow of AAAS. He received a BS degree in Electrical Engineering from APS University in Rewa, India in 1972 and an MS in Computer Science & Controls from IISc, Bangalore, India in 1974 and the Ph.D. degree in Applied Math/Computer Science from Harvard University in 1978. Dr. Jain is currently the Barbara J. and Jerome R. Cox Jr. Professor of Computer Science & Engineering at Washington University in St. Louis. Previously, he was one of the Co-founders of Nayna Networks, Inc - a next generation telecommunications systems company in San Jose, CA. He was a Senior Consulting Engineer at Digital Equipment Corporation in Littleton, Mass and then a professor of Computer and Information Sciences at Ohio State University in Columbus, Ohio. He has 14 patents and has written or edited 12 books, 16 book chapters, 65+ journal and magazine papers, and 10e5+ conference papers. He is a winner of the ACM SIGCOMM Test of Time Award, CDAC-ACCS Foundation Award 2009, and ranks among the top 90 in CiteseerX's list of Most Cited Authors in Computer Science.

Mohammed Samaka is an associate professor of Computer Science in the Department of Computer Science and Engineering (CSE), College of Engineering at Qatar University. He obtained his Ph.D. and Master from Loughborough University in England and a Post Graduate Diploma in Computing from Dundee University in Scotland. He obtained his Bachelor of Mathematics from Baghdad University. His current areas of research include Wireless Software Architecture and Technology, Mobile Applications and Services, Networking, e-Learning, and Computing Curricula Development.

This paper is a revised and expanded version of the paper, “Dynamic Analysis of Application Delivery Network for Leveraging Software Defined Infrastructures," presented at the IEEE International Workshop on Software Defined Systems, Tempe, AZ, USA, March 9-12, 2015, 2.

This work was made possible by the grant number NPRP 6-901-2-370 from Qatar National Research Fund (a member of Qatar Foundation). The statements made herein are solely the responsibility of the authors.

## 1 Introduction

Enterprises may obtain virtual resources from cloud service providers for their internal functions or to provide services to others. In the latter case, they would be known as application service providers (ASPs). Use of resources from single clouds has become commonplace in recent years in the government (Figliola and Fischer, 2015) and businesses (Aljabre, 2012). The enterprises are now turning to multiple public clouds, also known as intercloud (AlZain et al., 2013), for added benefits of lower cost, increased flexibility, greater reliability, reduced latency and a larger number of specialized features. On the flip-side, use of resources on multiple clouds brings in the complexity of not only interfacing with disparate clouds, but also the requirement of controlling the wide area network connecting the clouds.

The term software-defined infrastructures (SDI) refers to the virtualized resources that the Cloud Service Providers (CSPs) and the Network Service Providers (NSPs) offer through software-based control and management systems. The physical devices, on which these virtual infrastructures are created, could themselves be located in one or more datacenters of a cloud or diverse and geographically separated clouds each controlling one or more datacenters. Some examples of control and management platforms for individual clouds include OpenStack and EC2 that virtualize computing and storage resources and OpenDaylight doing the same for network resources. Software control of infrastructure allows the flexibility of application specific virtual clouds to be carved out of virtual resources from multiple clouds interconnected with virtual wide-area-networking resources and to control dynamically and manage them. SDIs provide ASPs with such a converged view of resources provided by the cloud service providers. In the case of multi-cloud applications, an ASP needs to see a converged view of resources across multiple clouds as if they were all on one cloud. This would allow them to use resources available from many providers, through their APIs, in a manner that enables optimization of flexibility, reliability, latency, capital and operational expenses.

The term application delivery network (ADN) has been used to refer to a distributed network architecture comprising of many different components, including 1) Application servers 2) Packet-level middleboxes 3) Message-level middleboxes, and 4) Network transport services, that are required to deploy and deliver modern applications. Applications deployed over multiple clouds share, in addition to compute and storage, networking resources as well. Virtualization creates isolated network contexts on the same physical infrastructure for tenants' application specific requirements. The Internet only gives best-effort performance and ASPs requiring performance guarantees can get it only through static pre-provisioning of resources or by creating smart overlays. A more expensive but surefire way for an enterprise is to have its own private network. However, for most enterprises adopting a multi-cloud strategy means that they would also need a shared network infrastructure that can satisfy their requirements. There is no standard interface through which applications can automatically and dynamically communicate their requirements to the network. OpenADN has been designed to solve this problem.

OpenADN is a multi-cloud control and management software that helps deploy applications employing, controlling and managing resources across multiple CSP clouds and NSPs' wide area networks. The general idea is simple: Large ASPs like Google have the resources to install application layer proxies at their points of presence in distributed locations so as to intercept service request and route it to the nearest datacenter. Through OpenADN smaller, network constrained ASPs can obtain such services from third party infrastructure providers, e.g., ISPs, who can route application messages through an appropriate set of controllers, proxies, and middleboxes. This way the ASPs can get the benefit of deploying distributed applications on multiple clouds to get increased responsiveness and resiliency economically [Paul et al., 2014].

Software presenting an integrated virtualized environment of physical resources distributed across a number of public clouds and operating under disparate control and management software tends to be a complex system. Modularity is important in such systems for ease of development, maintenance and fate decoupling of the processes. They generally use multithreading for concurrent execution of a number of activities. If the modules, of such a system, do not work in harmony, performance suffers resulting in inefficient resource utilization and greater energy consumption (Khan et al., 2011). In other words, if the platform software has not been optimized, then the resources would be inefficiently utilized, resulting in sub-optimal system behavior and increase in operational expenditure. Such systems also lead to higher energy consumption and are contradictory to the notion of reducing the carbon footprint. Enterprises may face such situations leading to performance degradation, during operation, in which case the combined cost of usage and maintenance is high. Alternatively, they may insist that the control and management software be optimized for their situation before commissioning. This would require understanding the behavior of the software, through profiling, if possible in the production environment. Using well-known techniques, software engineers can isolate hotspots (problem areas in the system) that consume a disproportionate share of resources leading to sub-optimal behavior. Multithreading adds another level of complexity. It makes the system difficult to profile because characterizing the effects of interactions between threads becomes difficult as described in Waddington (2009). Efficient abstractions need to be developed to capture this behavior without resulting in exponential analysis times.

Taking advantage of the first such system being available to us, we have attempted to characterize the behavior of OpenADN, under operation, and used several relevant profiling techniques to see what could cause the system to behave sub-optimally. This should spur the developers of such systems to fine-tune their platforms, saving money for the users and reducing energy consumption. Section 2 describes the OpenADN platform highlighting its distributed and multi-threaded nature. Section 3 deals with profiling approaches that can be used for platforms like OpenADN dealing with resources spread across multiple clouds. Section 4 takes up the discussion on experimental set-up and OpenADN profiling outcomes. It also discusses how these results could be useful in decisions about optimization. We draw conclusions from our study in Section 5.

## 2 Managing Software Defined Infrastructure Over Multiple Clouds – OpenADN

Most contemporary and future application deployments like Internet-of-Things (IOT), Cyber-Physical Systems, mobile apps, massively parallel gaming and virtual reality tend to be distributed and need to use multiple clouds primarily due to cost and latency considerations. ASPs can use OpenADN to manage such distributed applications as if they were deployed on a single cloud. In the following sections, we shall see why the architecture of OpenADN is suitable for such massively distributed application scenarios. It is also relevant to discuss these details as they affect the performance of the system.

### *2.1 Key Elements of OpenADN*

OpenADN is interposed between the clouds and the interconnecting network on one side and the application deployment environment on the other. It offers features for application architects, application developers, and application managers. As shown in Figure 1, on the north side, OpenADN offers interfaces for application personnel to define the application resource requirements and deployment policies. On the south side, it interacts with various clouds and wide area networks. The northbound interfaces of OpenStack/ OpenDaylight become southbound interfaces of OpenADN. OpenADN architecture has a modular structure similar to the OpenDaylight SDN controller [OpenDaylight, 2015] with many southbound interfaces. While OpenStack allows implementing client policies in one cloud, OpenADN allows implementation of client policies uniformly among all the clouds.

**Figure 1** OpenADN Multi-Cloud Management System

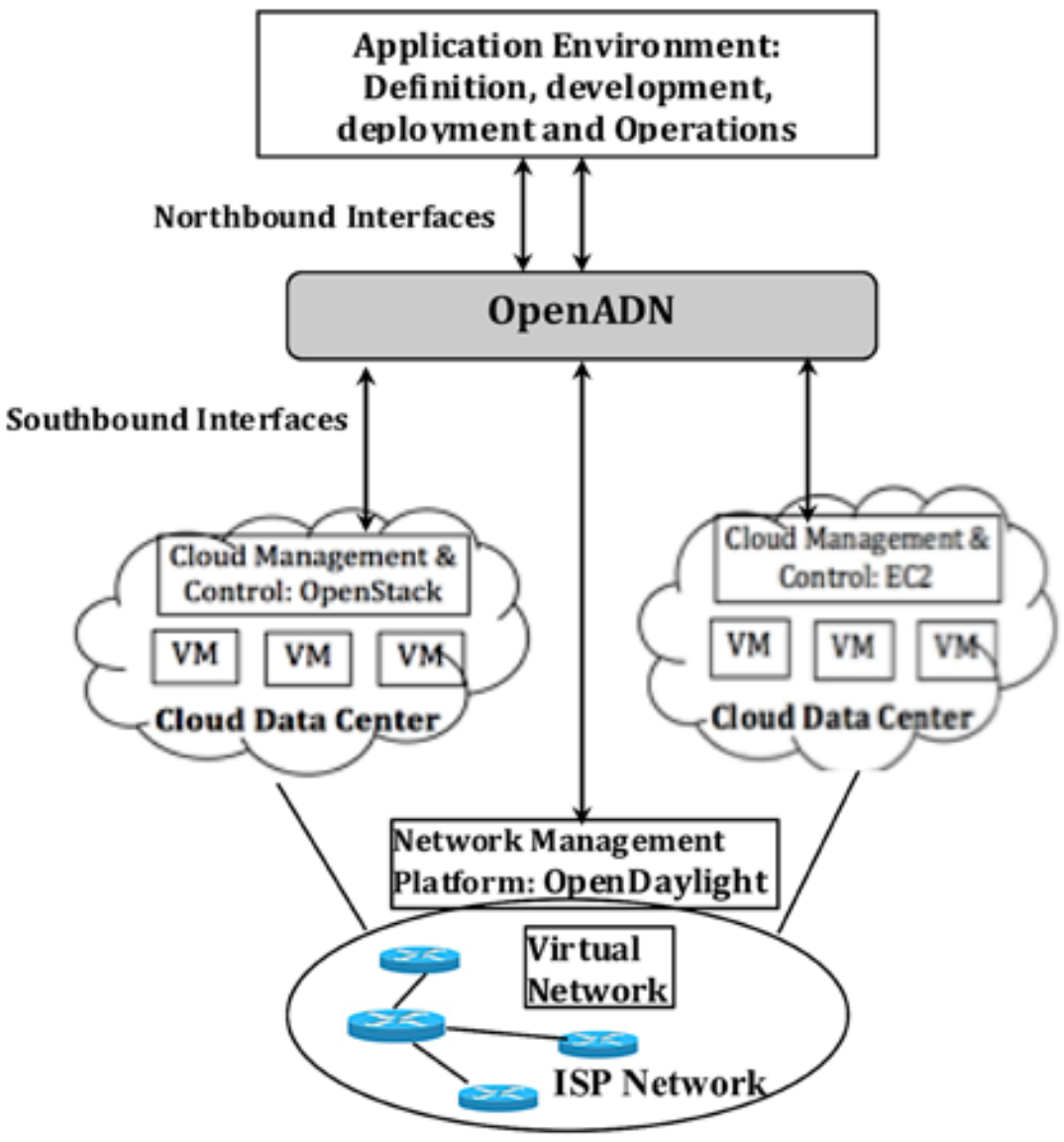

Instead of directly manipulating the resources inside the clouds, OpenADN simply requests the respective cloud manager to create those resources. As shown in Figure 2, the ASP specifies the policies regarding when and where to create the resources. The management plane is centralized while the control plan of OpenADN has a hierarchical architecture. A hybrid design has been chosen for the control plane with the centralized global controller and distributed local controllers to get the benefits of both designs. While distributed architectures are more scalable and also more resilient against failures and security threats, centralized architectures are simpler to manage. A proper division of work between the global and the local controllers ensures a good combination of latency and accuracy. The data plane is distributed to take advantage of distributed applications and the network. The key building blocks of OpenADN are shown in Figure 2

The global manager bootstraps the system and co-ordinates with the management platforms of various cloud providers for the acquisition of resources. Each of these resources, either owned or leased, is managed, controlled and programmed by the OpenADN control and management plane. The control plane is hierarchical, with a separate controller for each resource provider. A global controller, in turn, manages these local controllers, each residing on a virtual machine in the cloud. The control plane of OpenADN interacts with and programs the virtual resources like a virtual machine, virtual switch, and virtual router with the ASP's deployment and delivery policies through a southbound control interface.

**Figure 2** Key Building Blocks of OpenADN

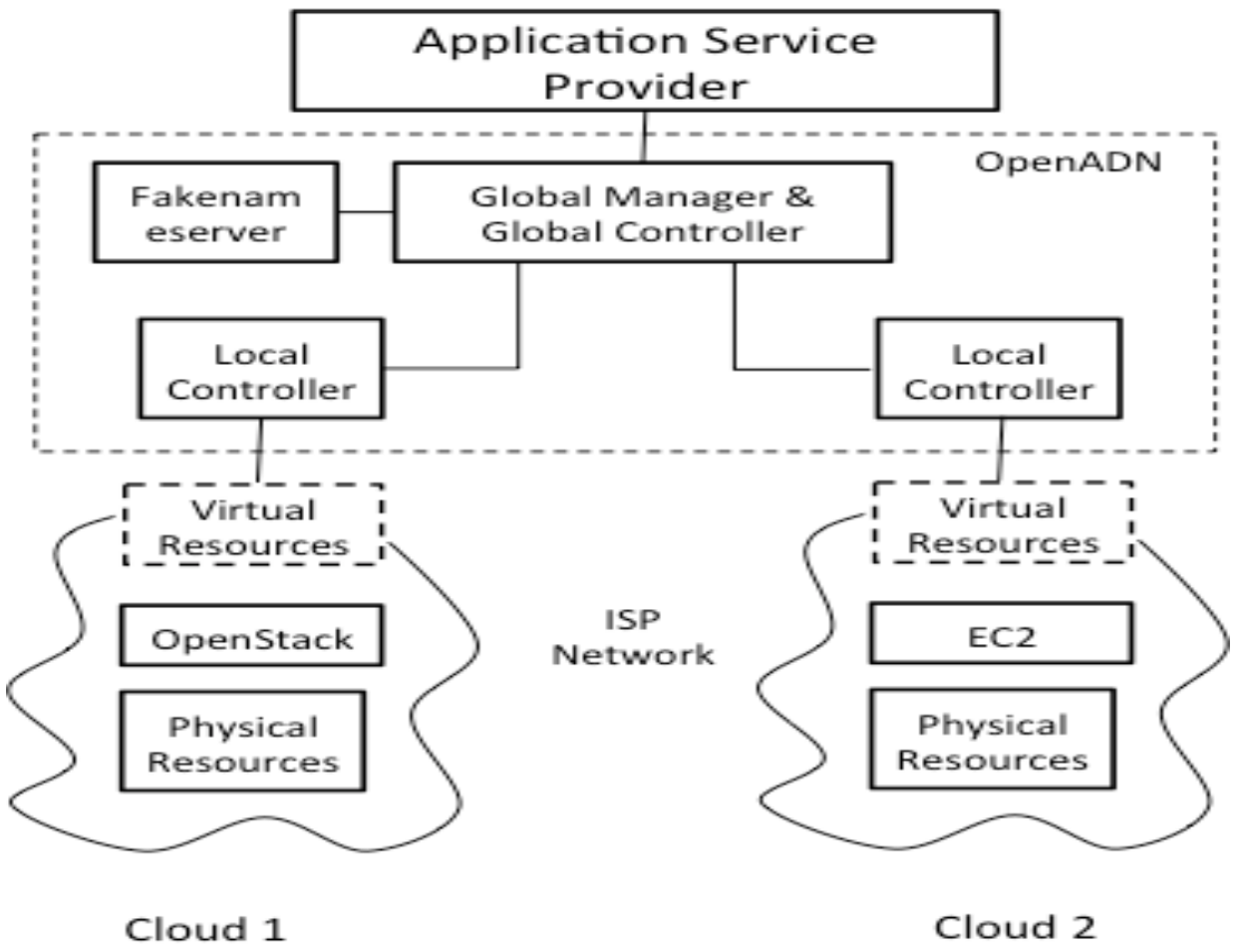

After bootstrap, the global controller takes over and launches one workflow manager for each workflow. The workflow manager checks for resources and launches workflow instances. The workflow manager commissions or decommissions workflows depending on the load. There is one local controller for each datacenter for quick local decisions. This works well with a highly distributed data plane. Considering the geographical spread of resources, the centralized global controller making it easy to introduce new services, propagate new policies and troubleshoot problems. OpenADN is an integrated infrastructure comprising both, message-level devices (e.g., firewalls) and packet-level devices (e.g., intrusion detection devices), hosting application-layer services as well as network-layer services. For massively distributed applications, like mobile healthcare monitoring or mobile app delivery, OpenADN allows multiple zones with each zone consisting of multiple clouds [Paul et al., 2014].

### *2.2 Massively Distributed Nature of OpenADN*

At the bootstrap stage, OpenADN creates a common global controller and one local controller for each data center. The global manager is manually started and, given the sensitive customer information it stores would normally be located in the ASP premises. Other functions can all be located on virtual machines in the cloud. OpenADN optimizes application service deployment by deploying the hosts of the distributed data plane, on virtual resources of various clouds. The system can perform many different tasks at the same time leading to better utilization of the hardware resources and ensure that the system as a whole makes progress all the time. OpenADN is essentially a multi-threaded system where performance is determined by the execution environment. This massively distributed data plane structure, with several threads in the state of operational stupor, makes the performance evaluation of OpenADN difficult and calls for specialized techniques that we shall discuss in the following sections.

## 3 Profiling Multi-cloud Delivery Platforms

In this section, we discuss the importance of profiling led optimization followed by a selection of techniques that are suitable for profiling OpenADN and gathering data for optimization.

### 3.1 *Profiling and optimization of modular multithreaded systems*

Often application software has code that consumes a disproportionate amount of resources and produces high CPU loads. Cloud management platforms are no different. The whole idea of profiling multi-cloud delivery platforms is to work through the tiers and threads of these platforms and collect information about their behavior in different operational situations. To this end, it is important to use program analysis tools that are appropriate to the distributed, multi-threaded nature of these platforms and gather as much information as possible. As against this, if we choose to carry out intuitive optimization, it may result in modification of parts of the code that were not responsible for performance degradation and as a result may be a waste of time. A word of caution: too much optimization or too little of it are both considered detrimental. Donald Knuth stated in [Knuth, 1974] that programmers waste an enormous amount of time thinking about the speed of non-critical part of the program. About 97% of the time we should forget about small inefficiencies as premature optimization is the root of all evil. It is not only important for profiling to precede optimization, someone who has tried to do it would realize that reading of such a code does not provide reliable information, and far less program behavior, under execution. It is, therefore, important to use correct techniques that would produce reliable data based on which it can be decided whether optimization should be carried out, and if the answer is in the affirmative, what parts of the code should be optimized (Eklov, 2012).

### 3.2 *Profiling Techniques for Software Platforms*

Given the nature of OpenADN, most conventional profiling, characterization, and modeling methodologies do not work well because of full system virtualization. They do not provide definitive help in pinpointing the sections of code that should be optimized. We shall see here a combination of techniques that can be applied to a distributed, multi-threaded system (Waddington et al., 2009). We divide these techniques into static, dynamic and concurrency profiling.

#### *1) Static Profiling*

In static analysis, program execution models are formally constructed (Jackson & Rinard, 2000). Models of multi-threaded systems can be used to explore all feasible inter-leavings and loops exhaustively to ensure correctness properties (Clarke et al., 2000). However, this kind of software is complex and may have a vast number of feasible inter-leavings making model checking computationally expensive. Another shortcoming of static-analysis techniques is that they give an assessment of relative time and temporal ordering and do not give absolute time (Rinard, 2001). For assessment of absolute times, it would be necessary to perform dynamic profiling (Mars & Hundt, 2009).

*2) Dynamic Profiling*

This type of profiling allows observing system behavior while it is running. In Waddington at al. (2009), it has been mentioned that dynamic profiling provides ways to measure the absolute time of events like various function calls or the time spent by the CPU in a particular function. It is an active form of profiling in which the system being measured explicitly generates information about its execution parameters. Conversely, passive profiling relies on explicit inspection of control flow and execution state through an external entity, such as a probe or modified runtime environment. Three main families of dynamic profiling techniques are code instrumentation, statistical sampling and concurrency profiling.

a) *Code Instrumentation*: A set of additional instructions called an instrument is injected into the target program. When the instrumented code is executed, it generates the required information. These instructions indicate events as they happen and provide their timing and frequency. Some instrumentation systems (Clarke et al., 2000) count function activations while others (Chen et al., 2010) count more fine-grained control flow transitions. This method can, thus, provide an absolute measure of these events. Instrumenting a program can cause changes in the performance of the program, potentially causing inaccurate results and has to be carried out carefully in a controlled manner.

b) *Statistical Profiling:* In this method the program state is randomly sampled when it is in execution. This involves recording a sample of values of the instruction register, program counter, stack, etc. to apply statistical techniques to these samples to deduce how much time is being spent in different parts of the program. This method is not as intrusive to the target program as the instrumentation method. They can show the relative amount of time spent in user mode versus interruptible kernel mode such as system call processing and also the user time out of the total execution time (Mars and Hundt, 2009 and Intel, 2007). In OpenADN environment, this could, for example, provide valuable information on whether optimization should at all be attempted.

c) *Deterministic profiling:* In this method all function calls, function returns, and exception events are monitored, and precise timings are obtained for the duration of these events and the intervals between them. OpenADN is largely written in Python. In Python, since there is an interpreter active during execution, the presence of instrumented code is not required to do deterministic profiling. Python automatically provides a *hook* (optional callback) for each event. Call count and time consumption statistics can be used to identify hotspots in code, which would be potential candidates for optimization.

3) *Concurrency Profiling*

Concurrency profiling can additionally be used in multithreaded applications. Resource contention profiling collects detailed call stack information every time that competing threads are forced to wait for access to a shared resource. Concurrency visualization also collects more general information about how a multithreaded application interacts with itself, the hardware, the operating system, and other processes on the hosts. It can help locate performance bottlenecks, CPU underutilization and synchronization delays (Microsoft, 2013 and Oracle, 2012)

## 4 Multi-Cloud Platform profiling

The complexity of the multi-threaded OpenADN platform necessitated collection of precise and fine-grained behavioral data while in execution, coupled with off-line analysis to help characterize the performance of the platform and possible need for optimization. Profiling of OpenADN was, therefore, carried out at multiple levels as shown in Figure 3. Platform level is a coarse grain profiling, which gives overall time spent in execution of the platform. Depending on the method used, this may give a broad idea about the time spent in useful activities and possibly in infructuous ones. The OpenADN module level profiling consists of profiling various functions while they interact with each other. This gives a fairly good idea of which system, external library or kernel functions are taking unduly long times. When required, we could do a still more fine-grain profiling at the statement level of the suspect functions to pin-point where the problem actually lies. We shall see how these were used in the case of OpenADN.

**Figure 3** Levels of OpenADN Profiling

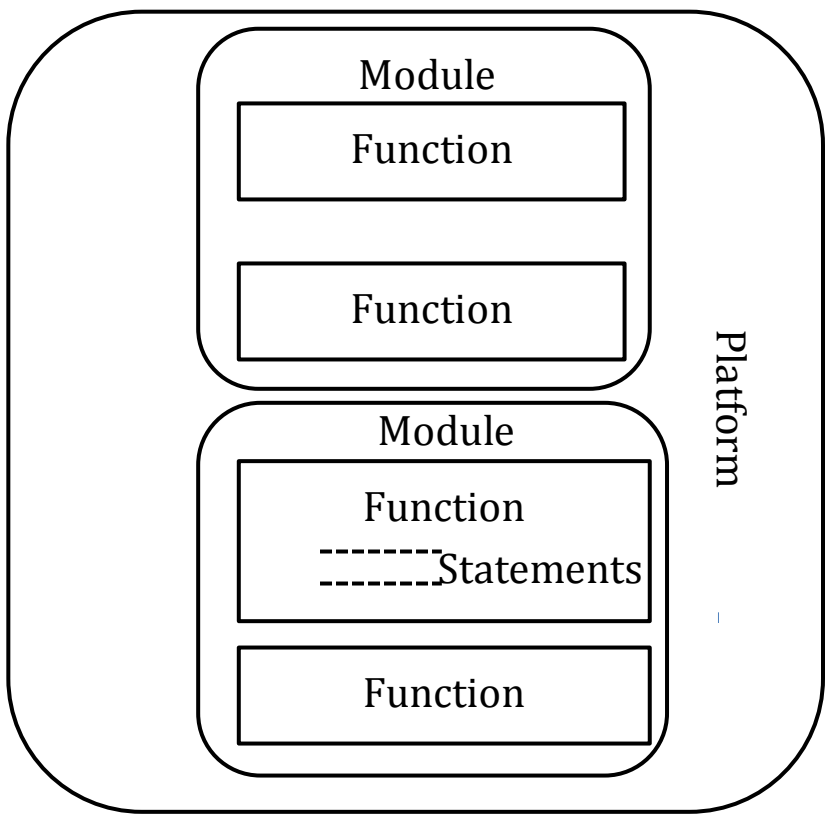

To validate the functionality, we ran OpenADN in a virtual environment created by Mininet [Lantz, 2015]]. Mininet allows emulating a whole virtual network running real kernel, switch and application code, on shared physical resources of a machine. The following virtual resources were created for profiling OpenADN: One service zone consisting of a global controller, two data center sites with a local controller each, a name-server, seven hosts per site and client host simulating 10,000 users. The selection of stimuli (set-up and input data) and multiple runs of the platform ensured that behavioral data for most control paths were collected.

To bring home the complexity of the OpenADN platform, we will briefly discuss its bootstrap and execution process using Figures 2 and 4 to explain the inter-relationship of the functional blocks.

**Figure 4** Functions of OpenADN relevant to the bootstrap process

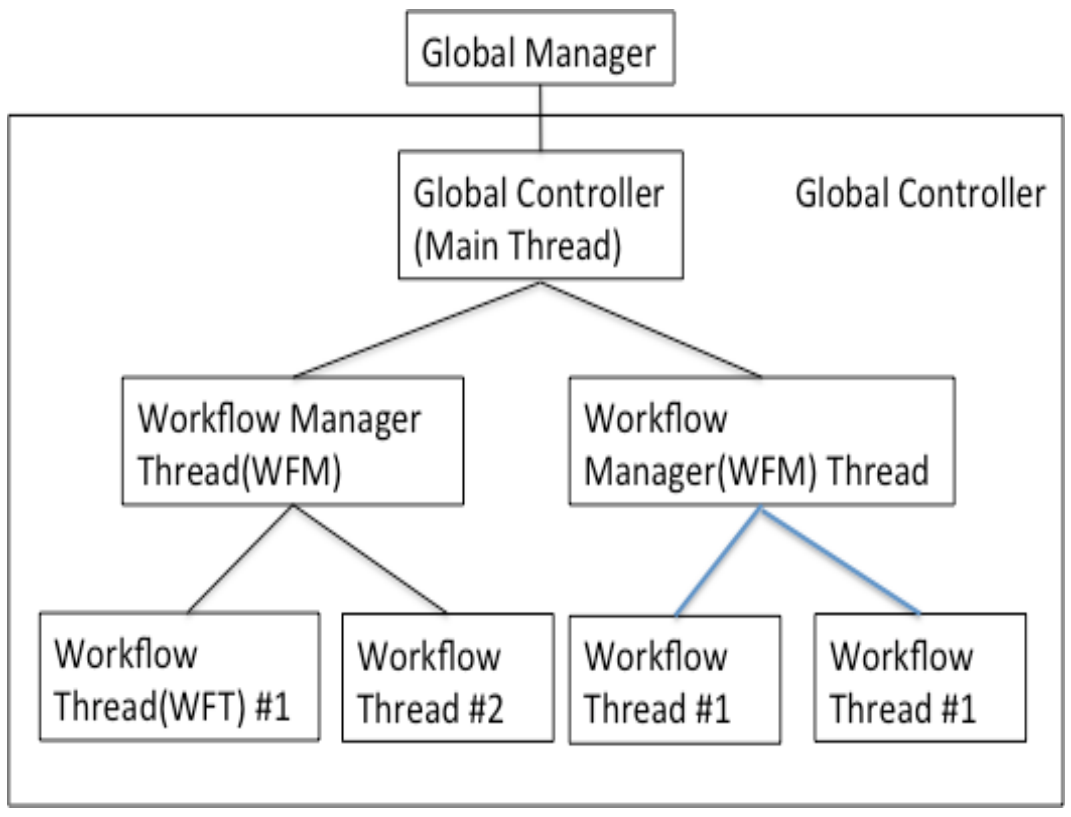

The global manager is the only module that needs to be started, and then the complete bootstrap process is automatic. The global controller launches one workflow manager for each zone. The flow of operations is as in Figure 5.

**Figure 5** The OpenADN bootstrap flow diagram

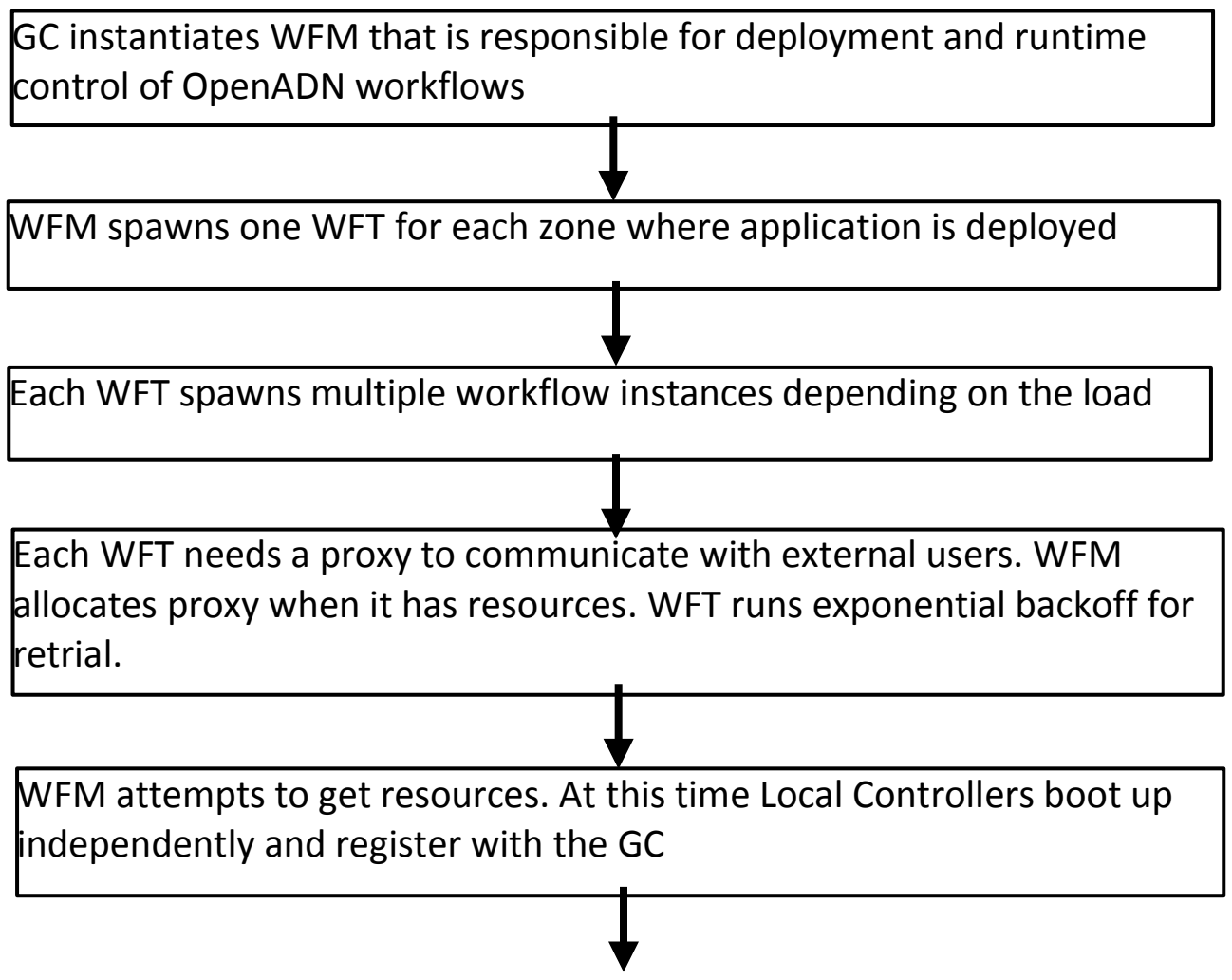

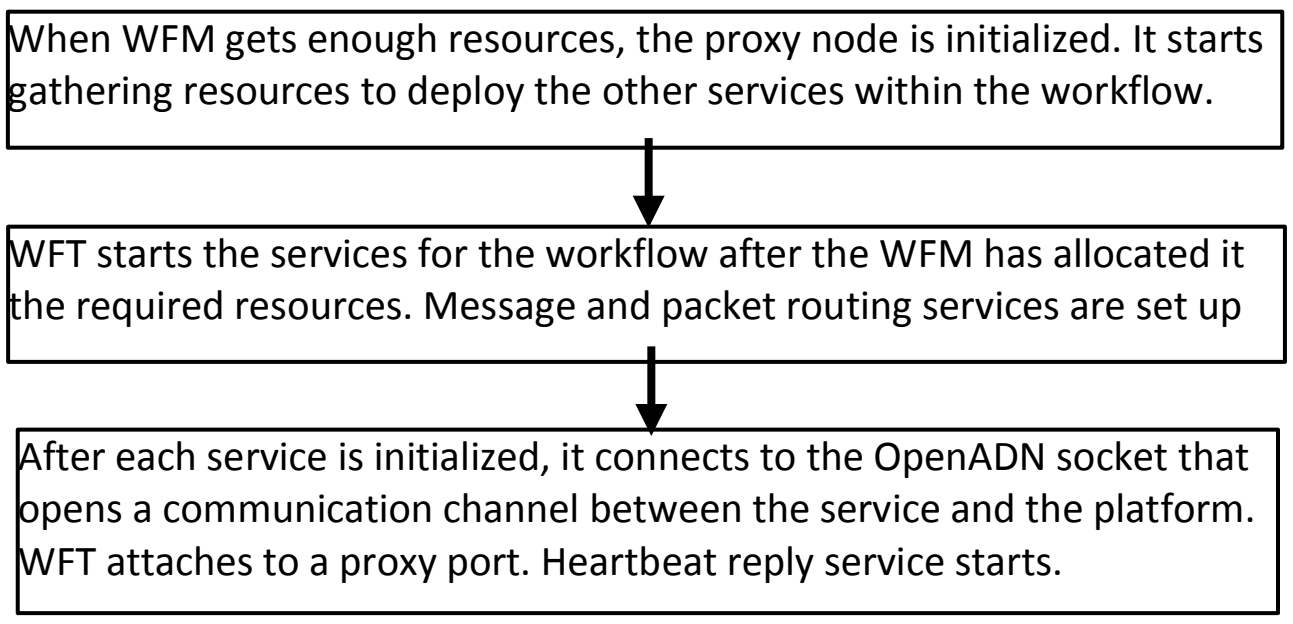

### *4.1 Coarse Grain Analysis*

To get a broad idea of the efficiency of the platform code executing in a virtualized environment, the Unix time utility was used. The platform software was run to create virtual hosts over which the platform modules – global controller, local controller, name server, node controller, and clients were loaded and executed. The program was run to bootstrap the process and until all the modules were added and services started running. Some runs were performed for the same virtual environment and data from five of them are given in Table I.

**Table 1** Run time used for user and system activities

| | | | | | | | | (Time unit: seconds) |
|---|---|---|---|---|---|---|---|---|
| **Run** | 1 | 2 | 3 | 4 | 5 | 6 | Average | % Run time |
| **User Space** | 0.62 | 0.59 | 0.55 | 0.63 | 0.72 | 0.74 | 0.64 | 1.45% |
| **System Calls** | 0.65 | 0.73 | 0.75 | 0.96 | 1.58 | 1.65 | 1.05 | 2.38% |
| **Run time** | 34.65 | 35.22 | 35.6 | 42.67 | 56.01 | 60.74 | 44.15 | 100 |

The elapsed time is the total platform run time for booting and starting new services, user-space time is for non-system calls or CPU time spent outside the kernel, and system-calls is time spent in kernel specific functions.

Of the average total elapsed time of 44.15 seconds for which the platform software was executed, the time spent in user functions and kernel space was 1.45% and 2.38%, respectively. This gives an indication that a large part of the total CPU time is spent in activities other than running user and system functions. It is possible that much of this time is being spent in I/O waits and sleep times for dealing with dependent asynchronous concurrent processes. However, it cannot yet be said whether this time relates to unavoidable delays and the situation can be improved through optimization. This called for the next level of profiling, i.e., at module/function level to see which of the modules are highly CPU intensive.

For comparison, the same modules were also run on separate physical machines for comparison and the results obtained are given in Table 2.

On physical machines, the platform does not have to spend time creating virtual machines for its own modules as well as for running services. Even in this case the overall user-space time is 17.49% and even less for kernel calls. Among these, the global controller used the time more effectively with user functions taking up to 41.64% of run time on an average. However, in the actual operational environment, these modules will be hosted on VMs that will take a finite amount of time to create, start, augment or migrate to another cloud.

**Table 2** Time used for user and system activities on physical machines

| **Function** | **User Space** | **System Calls** | **Run Time** | **User (%)** |
|---|---|---|---|---|
| **Name Server** | 14.161 | 5.072 | 229.438 | 6.17 |
| **Global Controller** | 83.637 | 15.797 | 200.835 | 41.64 |
| **Local Controller** | 18.549 | 7.16 | 175.57 | 10.57 |
| **Node Controller** | 19.95 | 8.86 | 156.99 | 12.71 |
| **Client** | 0.428 | 0.036 | 18.855 | 2.27 |
| Total | 136.725 | 36.925 | 781.688 | 17.49 |

This simple profiling indicates the possibility of higher load on the CPU, because of potentially wasteful activities like waiting on I/O calls and the sleep functions. While in many cases, where asynchronous linking of threads is used, some waiting would be unavoidable. However, one needs to see whether these could be optimized for 1) making the platform more efficient 2) correctly dimensioning the resources leased, and 3) distributing the workload properly.

### *4.2 Deterministic Profiling of Functions of OpenADN*

Deterministic profiling of OpenADN programs was carried out to see execution pattern and the resulting CPU loads of various functions. There are built-in profilers that provide information about how often and how long various functions execute. A profiler like ‘profile’ or ‘cProfile’ in combination with a function based on ‘*pstats*’ (Python, 2009-15) provide statistics to make that is amenable to analysis. Figure 6 gives a sample output.

**Figure 6** Sample deterministic profile run showing the creation of VMs and functions

```
Thu Apr 16 18:15:44 2015    crun1000.txt

         541337 function calls (541319 primitive calls) in 77.621 seconds

   Ordered by: cumulative time

   ncalls  tottime  percall  cumtime  percall filename:lineno(function)
        1    0.001    0.001   77.621   77.621 driver_mininet.py:2(<module>)
        1    0.849    0.849   77.613   77.613 driver_mininet.py:209(start_sim)
   111580   59.322    0.001   59.322    0.001 {built-in method poll}
        3   15.010    5.003   15.010    5.003 {time.sleep}
        1    0.001    0.001    5.183    5.183 driver_mininet.py:186(start_hosts)
        1    0.000    0.000    5.010    5.010 driver_mininet.py:163(start_fakeNameServer)
        1    0.000    0.000    5.009    5.009 driver_mininet.py:134(start_gc_lighthouseController)
     3106    0.013    0.000    1.065    0.000 driver_mininet.py:264(write)
      104    0.160    0.002    1.052    0.010 util.py:25(quietRun)
        1    0.000    0.000    0.980    0.980 driver_mininet.py:42(__init__)
        1    0.001    0.001    0.943    0.943 driver_mininet.py:67(allocate_singleSwitchTopo)
       19    0.001    0.000    0.893    0.047 node.py:300(linkTo)
       19    0.001    0.000    0.698    0.037 util.py:79(makeIntfPair)
     3106    0.671    0.000    0.671    0.000 {method 'write' of 'file' objects}
     3106    0.381    0.000    0.381    0.000 {method 'flush' of 'file' objects}
      195    0.003    0.000    0.294    0.002 node.py:235(cmd)
    40024    0.287    0.000    0.287    0.000 {time.time}
      125    0.004    0.000    0.273    0.002 subprocess.py:619(__init__)
        1    0.000    0.000    0.266    0.266 driver_mininet.py:128(start_topo)
```

This run has been ordered on cumulative time (cumtime), which is the total CPU time (in seconds) a function executes including all the sub-functions it calls. The total number of calls to a function is in column ‘ncalls’ while the ‘tottime’ gives the CPU time excluding time taken by sub functions. The first ‘percall’ column does not include CPU time taken by sub-functions while the second ‘percall’ column includes those. In this sample run, it can be seen that the total time that the platform software was executed was 77.621 seconds. Out of this, the simulator module took 77.613 seconds. Creation of virtual nameServer (called Fakenameserver) takes 5.010 seconds while global controller takes 5.009 seconds.

Another section of the simulation is shown in Figure 7. Creation of virtual resources and starting of functions like global controller, local controller, etc. was observed for several runs and these operations took an average of about 18.5% over the three runs shown in Table 3. It was initially suspected that the ZeroMQ messaging library [12] poller takes up a lot of CPU time. This was confirmed by a number of runs shown in Table 3. It can be calculated that polling operation took on an average 79.83% of the CPU time over these runs.

**Figure 7** Sample run showing creation of local controller and client_host

```
         19    0.000    0.000    0.033    0.002 net.py:153(addHost)
         19    0.000    0.000    0.032    0.002 node.py:348(setIP)
          1    0.000    0.000    0.031    0.031 node.py:585(start)
          1    0.000    0.000    0.031    0.031 node.py:578(setup)
76711/76705    0.029    0.000    0.029    0.000 {len}
          2    0.000    0.000    0.022    0.011 moduledeps.py:60(pathCheck)
        125    0.020    0.000    0.020    0.000 {posix.fork}
          1    0.000    0.000    0.016    0.016 moduledeps.py:25(moduleDeps)
          2    0.000    0.000    0.016    0.008 moduledeps.py:7(lsmod)
          1    0.000    0.000    0.012    0.012 driver_mininet.py:175(start_client_host)
          1    0.000    0.000    0.008    0.008 driver_mininet.py:143(start_lc_lighthouse
        172    0.000    0.000    0.008    0.000 util.py:53(isShellBuiltin)
          1    0.000    0.000    0.008    0.008 node.py:634(start)
          3    0.000    0.000    0.007    0.002 node.py:450(sendCmd)
          1    0.000    0.000    0.006    0.006 net.py:560(init)
          1    0.001    0.001    0.006    0.006 util.py:1(<module>)
          1    0.000    0.000    0.006    0.006 net.py:184(addController)
```

**Table 3** CPU Time of selected functions

| Run Time | VM Creation | VM Starting | VM Creation and Starting% | Poll | Poll % |
|---|---|---|---|---|---|
| 100.969 | 15.441 | 2.162 | 17.43 | 81.948 | 81.16 |
| 77.621 | 15.255 | 2.155 | 22.43 | 59.687 | 76.90 |
| 105.849 | 15.482 | 2.22 | 16.72 | 85.439 | 80.72 |

This shows that there is a possibility that these operations can be studied further and optimized. However, at this stage, we are not sure which parts of the functions need to be looked into for higher consumption of CPU time. Polling library function, for instance, is called a number of times in many user functions. It would, therefore, be of interest to see how different statements within each function execute.

If we dissect the virtual machine creation time among different functions, we see that the time taken by some of the important ones, for a typical run, as given in Table 4.

**Table 4** Time for different functional modules

| Module | CPU Time (seconds) |
|---|---|
| Fakenameserver | 5.010 |
| Global controller | 5.009 |
| Local controller | 0.008 |
| Hosts | 5.183 |
| Client host | 0.012 |

We will see later that the Name Server and the Global Controller sleep through most of the time. Their job is largely reactive in nature, getting activated when other modules need their services.

The sleep function is required to be invoked at several places in multi-threaded software to allow interacting processes to wait for the required input to be available from the other processes. These times are sometimes decided based on intuition and contribute towards increasing total run time.

### *4.3 Statement Level Analysis*

Profiling at the platform and function levels provided a good idea of the time spent by one or more CPUs in kernel space calls, user space calls and waiting for I/O and in various modules of OpenADN. It was observed that a large proportion (96.17%) time was spent in waiting for I/O. Polling operations took about 79% of the execution time. The program spent about 15 seconds of the total 100 seconds in sleep mode. OpenADN functional modules took up to about 5 seconds each. Translated over long operational periods some of these have the potential to become the antithesis to the efficient operation of the platform.

The function level profiling, in particular, yielded CPU times for functions related to creation and loading platform modules, obtaining the resources and bootstrapping the platform. In this case, it confirmed that certain functions, e.g., the creation of virtual machines and linking, polling, heartbeat operation and sleep functions take unduly long part of the run time. The information from function-level profiling was not enough to tell us which modules to look into to locate the potential hot spots and optimize the software. As is often the case, the reason for a particular module or functionality taking a large amount of time could be pinpointed to some small part which may seem to be innocuous on a simple reading of the code. Some statements could trigger a library function or call to a special method that may not be so obvious. The function level profiling only times the explicit function calls and not the special methods called. Such profiling would not identify a slow operation in the library function like ZeroMQ. If a statement triggers the computation when using libraries, when there is no explicit call, function profiler will not go into the constituents.

A more detailed statement level analysis of the platform software was undertaken to determine which parts of the program take more CPU time. A more intrusive line profiler that could go into each function and time execution of each statement was used for this purpose. The line-profiler described in Kernprof (2015) used in a judicious manner allows this kind of analysis. This profiler keeps track of multiple statement executions, sums up the total time each statement takes in multiple passes and avoids profiling overheads. The profiling result is a binary file that could be deciphered with 'pstats' or a similar function. The output consists of the following:

a) Hits: Number of times that line was executed.
b) Time: Total execution time
c) Per Hit: Average amount of execution time
d) % Time: Percentage of time spent on that line relative to the total amount of recorded time spent in the function.
e) Line Contents: Actual source code.

The illustrations in Figure 7 to 11 show some of the portions of profiling data that indicate a possible need for optimization (column headings of figure 7 apply to figures up to 11). Figure 8 and nine show linking to the switch takes up a major percentage of the execution time. The name-server takes 9.9%, while the global controller takes 4.5%. The hosts take the longest accounting for 53% of the time.

**Figure 8** Creation of topology

```
Timer unit: 1e-06 s

Total time: 0.62231 s
File: driver_mininet.py
Function: allocate_singleSwitchTopo at line 69

Line #      Hits         Time  Per Hit   % Time  Line Contents
==============================================================
    69                                           @profile
    70                                           def allocate_singleSwitchTopo(self):
    71         1           13     13.0      0.0          yappi.start(builtins=False, profile_threads=True)
    72                                                   #pr=cProfile.Profile() #remove
    73                                                   #pr.enable() #remove
    74         1         2472   2472.0      0.4          self.switch = self.net.addSwitch('s1')
    75         1         6223   6223.0      1.0          self.net.addController( 'c0' )
    76
    77                                                   # Fake name server
    78         1         5003   5003.0      0.8          self.gc_fakeNameServer = self.net.addHost("FakeNS", mac=self.ns_mac,
    79         1        56369  56369.0      9.1          self.gc_fakeNameServer.linkTo(self.switch)
    80         1          408    408.0      0.1          print ("adding global nameserver<%s, %s, %s>"%(self.gc_fakeNameServer
    81
```

**Figure 9** Creation of global controller

```
78   1                                #global controller
79   1       47       47.0     0.0    mac= util.macColonHex(1)
80   1       12       12.0     0.0    ip=util.ipStr(util.ipNum(10,10,0,1))
81   1   214515   214515.0     0.3    self.gc_lighthouseController = self.net.addHost("gc" , mac=mac, ip=ip)
82   1  3405763  3405763.0     4.2    self.gc_lighthouseController.linkTo(self.switch)
83   1    36048    36048.0     0.0    print ("adding flobal lighthouse controllr <%s, %s, %s>"%(self.gc_lighthouseController.name, mac, ip))
```

Figure 10 indicates that the linking of hosts to the network takes about 51.6% of the time.

**Figure 10** Profile run for linking hosts to a switch

```
121
122     6        21     3.5    0.0                          hostName = "h" + str(j-1)+ "s"+ str(i)
123     6      2362   393.7    0.4                          print ("adding host to site %s: <%s, %s, %s>" %(i, host
124     6      8833  1472.2    1.4                          host = self.net.addHost(hostName,mac=mac, ip=ip)
125     6    321058 53509.7   51.6                          host.linkTo( self.switch )
126                                                         # store info for each host
127     6        25     4.2    0.0                          hostInfo = {}
128     6         8     1.3    0.0                          hostInfo["host"] = host
129     6        24     4.0    0.0                          hostInfo["lc_controller_addr"]= lc_lighthouseControlle
130     6        11     1.8    0.0                          hostInfo["host_addr"] = host.defaultIP
131
132     6        35     5.8    0.0                          siteDesc[i-1]["hostList"].append(hostInfo)
133                                                 #       self.hostList[i].append (hostInfo)
134     2         8     4.0    0.0              self.siteDescList.append(siteDesc[i-1])
```

Figures 11 and 12 indicate that a large amount of time is taken up by the sleep function (the duration of which is sometimes programmed arbitrarily for synchronization) and the polling function during different phases of simulation. The global controller sleeps most of its execution time, and similar is the case with the name-server. This could mean that these functions are demanding more virtual resources than necessary and are leading to higher operational expenditure.

**Figure 11** Time spent in sleep function in the global controller module

```
Total time: 5.08587 s
File: driver_mininet.py
Function: start_gc_lighthouseController at line 149

Line #      Hits         Time  Per Hit   % Time  Line Contents
==============================================================
   149                                           @profile
   150                                           def start_gc_lighthouseController(self):
   151         1           16     16.0      0.0          print ("2. Starting Global Lighthouse Controller"),
   152         1          258    258.0      0.0          self.gc_lighthouseController.cmd('export HOST_NAME=%s'%(self.gc_ligh
   153         1          182    182.0      0.0          self.gc_lighthouseController.cmd('export NAME_SERVER_ADDR=%s'%(self.
   154         1          179    179.0      0.0          self.gc_lighthouseController.cmd('export NAME_SERVER_UPDATE_PORT=%s'
   155         1         6597   6597.0      0.1          self.gc_lighthouseController.cmd("python3 %s &"%(GLOBAL_CONTROLLER))
   156         1          579    579.0      0.0          print (".... started\n")
   157
   158
   159         1      5002929 5002929.0    98.4          sleep(5)
   160         1        74569  74569.0      1.5          yappi.get_func_stats().print_all()
   161         1          557    557.0      0.0          yappi.get_thread_stats().print_all()
```

**Figure 12** Time taken by polling function

```
271                                              #start the processes in the hosts
272        1        854     854.0      0.0       print ("-------------\n")
273        1    5050107 5050107.0       2.3      simNetwork.start_fakeNameServer()
274        1        129     129.0      0.0       print ("checkpoint 1...after fakenameserver")
275        1    5085906 5085906.0       2.3      simNetwork.start_gc_lighthouseController()
276        1        126     126.0      0.0       print ("checkpoint 2 ...after gc")
277        1     330151  330151.0      0.1       simNetwork.start_lc_lighthouseControllers()
278        1       2657    2657.0      0.0       print ("checkpoint 3 ..after lc")
279        1        229     229.0      0.0       print ("\n")
280        1        213     213.0      0.0       print ("4. Starting hosts:")
281        1        286     286.0      0.0       print ("checkpoint 4...after hosts")
282        1    5465310 5465310.0       2.5      simNetwork.start_hosts()
283
284        1     468482  468482.0      0.2       simNetwork.start_client_host()
285        1       4453    4453.0      0.0       print ("-------------\n")
286        1        268     268.0      0.0       print ("checkpoint 5...after client host")
287                                              #start the monitoring
288        1          8       8.0      0.0       endTime = time() + _runTime
289   140057    1011019       7.2      0.5       while time()< endTime:
290   140057  200834072    1433.9     90.5               readable = poller.poll(1)
291   148007     666242       4.5      0.3               for fd, _mask in readable:
292     7951      26630       3.3      0.0                       node = Node.outToNode[ fd ]
```

Also, the function to check the ports for inter-process messages takes up 90.5% of the entire simulation time.

## 4.4 Concurrency Profiling of OpenADN

While the recursive function level profiling that includes timing of execution of sub-functions and statement level profiling includes the effect of execution of various threads, the interaction may not be evident. To get a better understanding of the multi-threaded platform, a thread-aware profiling was carried out while the program was in execution. A sample of concurrency profile is given in Figure 13. This aspect of profiling is part of the future work.

**Figure 13** Concurrency profiling output

```
2. Starting Global Lighthouse Controller .... started


Clock type: CPU
Ordered by: totaltime, desc

name                                  ncall  tsub      ttot      tavg
...7.egg/mininet/util.py:25 quietRun  179    0.372231  0.983090  0.005492
..gg/mininet/node.py:300 Host.linkTo  35     0.001244  0.770000  0.022000
..gg/mininet/util.py:79 makeIntfPair  35     0.001611  0.541088  0.015460
..on2.7/subprocess.py:757 Popen.poll  65625  0.085931  0.333069  0.000005
..ckages/line_profiler.py:95 wrapper  3/2    0.000031  0.278079  0.092693
..et.py:141 mininetDriver.start_topo  1      0.000065  0.248423  0.248423
..g/mininet/net.py:348 Mininet.start  1      0.000077  0.248282  0.248282
..ocess.py:1256 Popen._internal_poll  65804  0.152177  0.247391  0.000004
..g/mininet/net.py:303 Mininet.build  1      0.000024  0.228708  0.228708
..net/net.py:255 Mininet.configHosts  1      0.001468  0.228668  0.228668
..g/mininet/node.py:267 Host.addIntf  70     0.000413  0.226883  0.003241
..7.egg/mininet/util.py:120 moveIntf  35     0.000189  0.226470  0.006471
..py2.7.egg/mininet/util.py:91 retry  35     0.000211  0.226280  0.006465
..ininet/util.py:105 moveIntfNoRetry  35     0.001316  0.226069  0.006459
..7/subprocess.py:619 Popen.__init__  216    0.008887  0.090917  0.000421
..ocess.py:1099 Popen._execute_child  216    0.031574  0.074085  0.000343
..7.egg/mininet/node.py:235 Host.cmd  153    0.003151  0.029812  0.000195
```

The number of times a function is called is given by ‘ncall’ while ‘tsub’ is the time spent in a given function excluding subfunctions. The CPU time of functions including its subfunctions is ‘ttot’. The average time ‘tavg’ is the average time spent in a function and its subfunctions in each call.

This type of data could be used in conjunction with statement level profiling to see the times of execution of various statements vis-a-vis the threads to which they belong. Total time the CPU spends in a thread can be derived from the information available. In this paper, we base our conclusions on the profiling levels shown in Figure 3.

## 5 Conclusions And Future Work

Multi-cloud management systems can be quite complex and may have parts of the code that may consume a big share of CPU time during execution. This could lead to suboptimal application delivery, increased resource usage, and higher operational expenses. ASPs who use such systems would like to optimize their platforms to control their expenses over their operational lifetime and for other desirable features like reduced latency and reduced energy consumption.

Intuition and reading of such multi-threaded code may not provide reliable information about what could be wrong with it. It becomes necessary to generate program profiles with data collected at various levels, i.e., platform, functions, and statements.

For OpenADN, the top-level analysis reveals that the overall execution time has a large component of non-user, non-kernel time that could be explained by I/O waits. A concern that arises is that some part of this time could be spent unproductively using up resources and increasing the cost. If this is actually the case then OpenADN, or any such multi-cloud management platform, would not be able to optimize the use of virtual resources resulting in the instantiation of a larger than the required number of virtual machines because the ones that have already been started are unproductively busy. Similarly the un-optimized use of virtual network resources would result in a higher payout to the network infrastructure provider. A function level analysis makes apparent the functions that have potential hotspots. Design choices at the time of the development of OpenADN govern the use of functions that might cumulatively consume substantial time. Non-blocking input-output in the form of polling or putting processes to sleep are an example of these. We have observed that the processes of such a platform might slow down if the use of these functions is not optimized. Statement level profiling on all of the modules simultaneously allows interplay of threads and reveals the parts of the functions that could be helped with optimization efforts. In the case of OpenADN, it was confirmed that asynchronous input-output with the use of polling, sleep functions and heartbeat were consuming a disproportionate amount of processing resources. If they were not optimized carefully, the resulting inefficiency would have resulted in higher operational expenditure for the ASPs.

Various runs of the platform with increasingly fine-grain profiling produced a large amount of data that was useful in deciding whether optimization should be attempted and, if it should, then what parts need to be tackled. Optimization is expensive and needs to be based on proper analysis of profiling data. From the data collected for optimization, it was found that creation and interlinking of hosts to the network, polling of ports for inter-service communication, heartbeat used to check aliveness of the services and use of sleep functions accounted for a large amount of time and could be potential candidates for optimization efforts.

Optimization could simply mean fine-tuning the sleep/wait times of processes built into the platform. On the other hand, there could be more serious issues, and optimization could involve a rewriting of some parts of the code.

Based on the profiling data, optimization could involve the following:

1. Critically examining the time spent in I/O waits (i.e. repeated polling of sockets for inter-service communication and request response mechanisms) and taking remedial measures wherever possible.
2. Examining the use of sleep statements and fine-tuning their durations.
3. Examining the use of heartbeat and ways to make it efficient.
4. Optimizing the time take to create dynamically, destroy and migrate virtual resources.

Profile-led optimization makes use of the results generated by deterministic, functional and statement level profiling to get optimized code. If the execution environment fairly represents the usage scenario, then profile guided feedback benefits optimization. Future work will involve demonstrating the usefulness of the approach in carrying out optimization of OpenADN. The scope of the problem at hand, however, was to see whether a combination of carefully selected profiling tools, working at different levels of the OpenADN program hierarchy (and by extension other similar platforms), would be able to pinpoint the bottlenecks that could cause higher consumption of virtual resources. From the results discussed above, it is clear that it would be in the interest of reduced cost and increased agility of doing the ASP business to carry out appropriate profiling at different levels as a precursor to optimization.

## References

Aljabre, A. (2012) "Cloud Computing for Increased Business Value," International Journal of Business and Social Science Vol. 3 No. 1

AlZain, M.A., Soh, B., Pardede E. (2013) "A Survey on Data Security Issues in Cloud," Journal Of Software, Vol. 8, No. 5

Chen, D., Vachharajani, N., and Hundt R. (2010) ''Taming Hardware Event Samples for FDO Compilation, Proc. 8th Ann. IEEE/ ACM Int'l Symp. Code Generation and Optimization (CGO 10), ACM Press, pp. 42-52.

Clarke, E.M., Grumberg, O. and Peled D.A. (2000) *Model Checking.* The MIT Press, Massachusetts Institute of Technology, Cambridge, Massachusetts.

Eklov, D., Nikoleris N. and Hagersten E. (2012) "A Profiling Method for Analyzing Scalability Bottlenecks on Multicores," ACM.

Figliola, P.M., Fischer, E.A. (2015) "Overview and Issues for Implementation of the Federal Cloud Computing Initiative: Implications for Federal Information Technology Reform Management," Congressional research service, CRS Report (7-5700)

Intel Whitepaper, “Optimizing Software for Multi-core Processors,” http://www.intel.com/content/www/us/en/intelligent-systems/intel-technology/multicore-optimizing-software.html

Jackson, D. and Rinard, M. (2000) “Software Analysis: A Roadmap,” *Proceedings of the IEEE International Conference on Software Engineering*, pp. 133-145.

Kernprof (2015) “Kernprof Line_Profiler,” https://github.com/rkern/line_profiler.

Khan, M.A., Hankendi, C., Coskun, A.K. and Herbordt, M.C. (2011) “Software Optimization for Performance, Energy, and Thermal Distribution: Initial Case Studies,” *International Green Computing Conference and Workshops (IGCC)*, pp. 1-6.

Knuth, D., “Structured Programming with go to Statements,” Computing Surveys, vol 6, No 4, December 1974.

Lantz, B., Handigol, N., Heller, B. and Jeyakumar, V. (2015) “Introduction to Mininet,” https://github.com/mininet/mininet/wiki/Introduction-to-Mininet.

Rinard, M. (2001) “Analysis of Multithreaded Programs,” *Proceedings of the 8th International Symposium on Static Analysis*, pp. 1-19.

Mars, J. and Hundt, R. (2009) ‘‘Scenario Based Optimization: A Framework for Statically Enabling Online Optimizations,’’ *Proc. 2009 Int’l Symp. Code Generation and Optimization (CGO 09)*, IEEE CS Press, pp. 169-179.

Microsoft (2013) “Concurrency Profiling,” http://msdn.microsoft.com/en-us/library/dd264994.aspx

OpenDaylight, http://www.opendaylight.org/project/technical-overview

Oracle (2012) “Multithreaded Programming Guide,” “Timers, Alarms, and Profiling,” https://docs.oracle.com/cd/E26502_01/html/E35303/gen-90808.html.

Paul S., Jain R., Samaka M. and Pan J., “Application Delivery in Multi-Cloud Environments using Software Defined Networking,” *Computer Networks Special Issue on cloud networking and communications*, Feb 2014

Hintjen, P. (2014) “The ZeroMQ Guide,” zguide.zeromq.org/page:all

Python (2015) “The Python Profilers,” https://docs.python.org/2/library/profile.html

Waddington, G.D., Roy N. and Schmidt D.C. (2009) “Dynamic Analysis and Profiling of Multi-threaded Systems,” IGI Global.

Figures, images and tables.

**Figure 1** OpenADN Multi-Cloud Management System

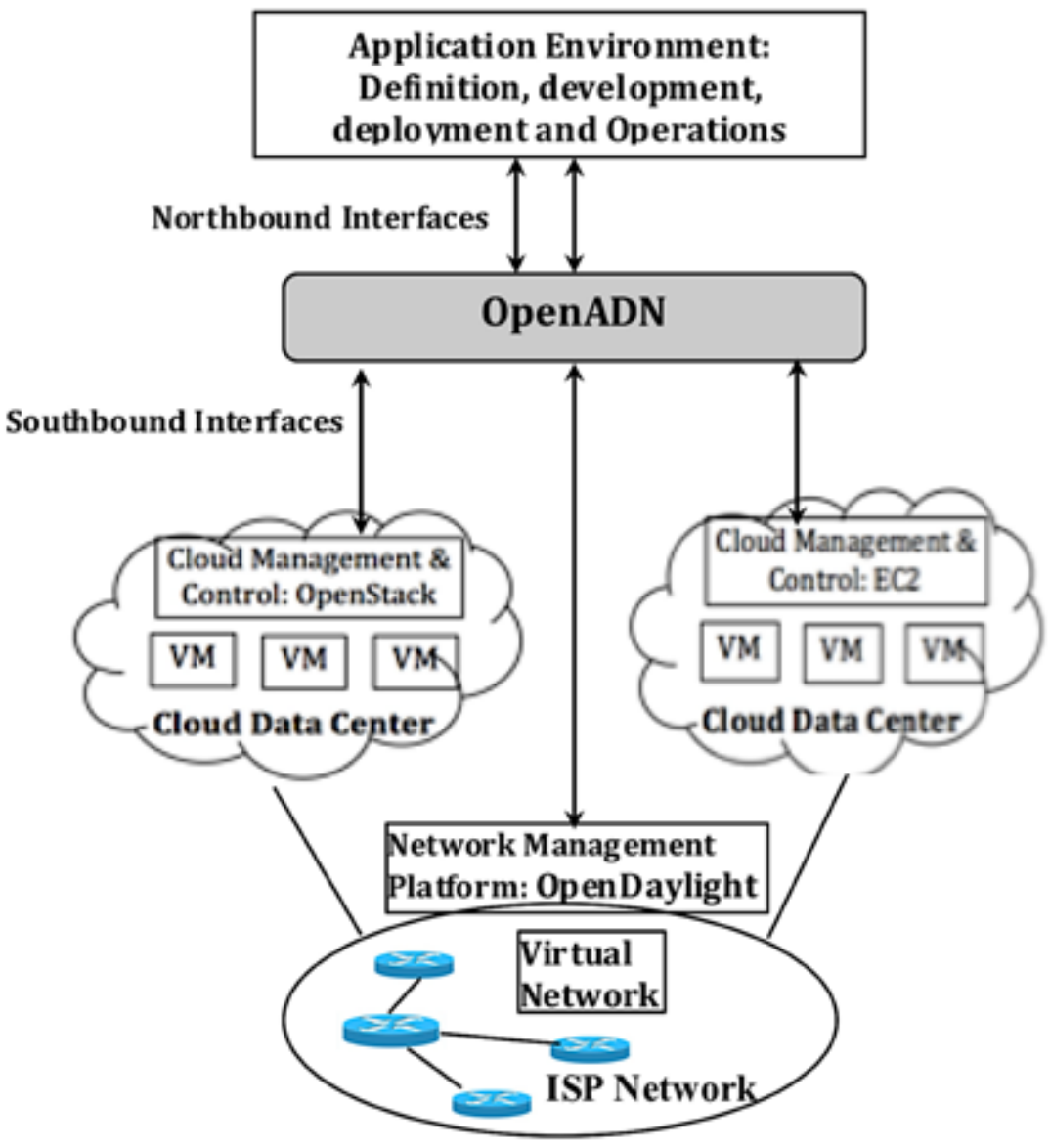

**Figure 2** Key Building Blocks of OpenADN

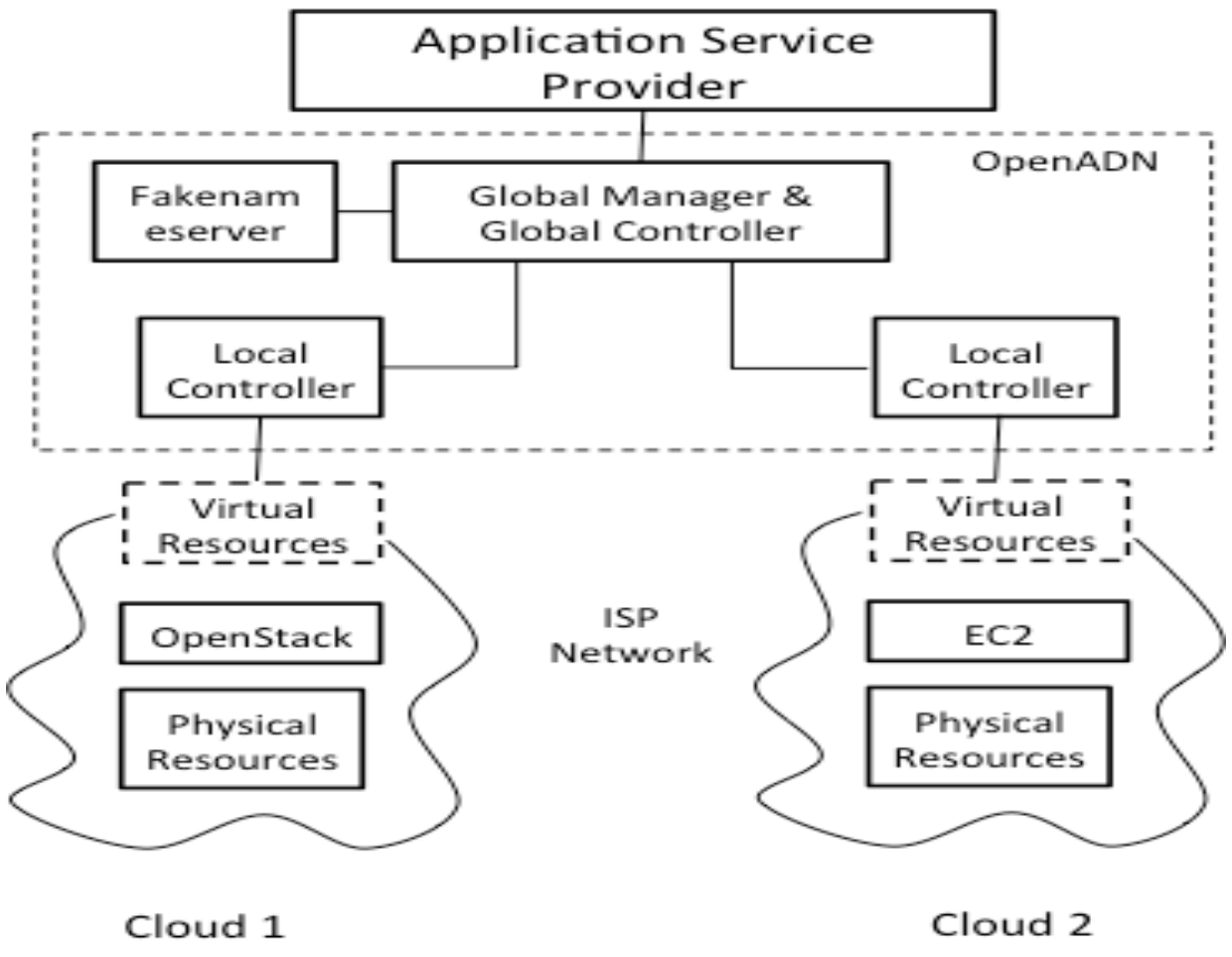

**Figure 3** Levels of OpenADN Profiling

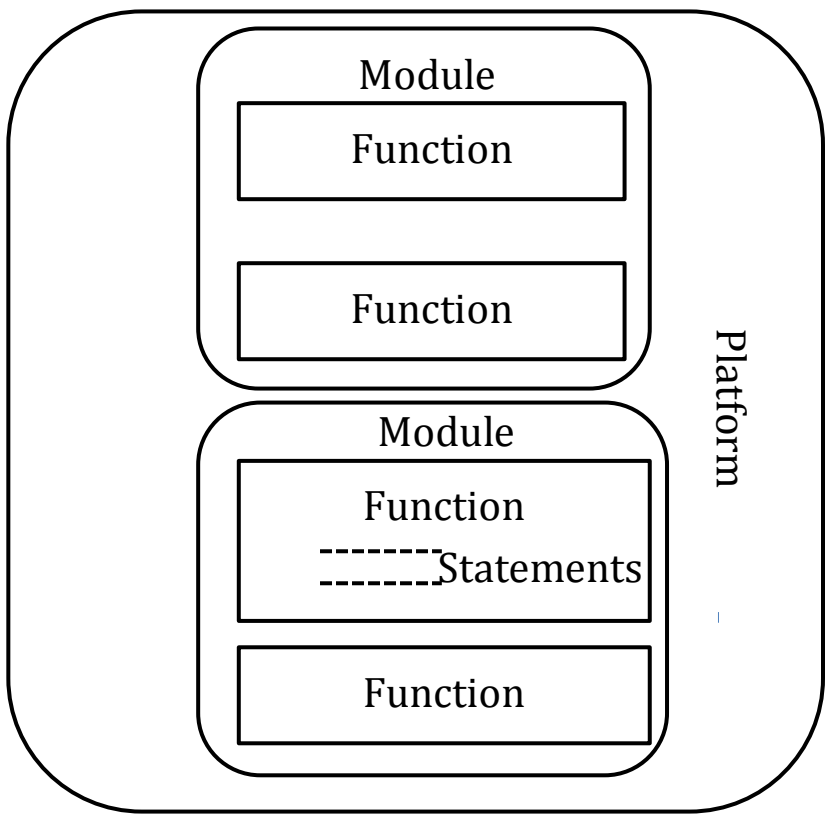

**Figure 4** Functions of OpenADN relevant to the bootstrap process

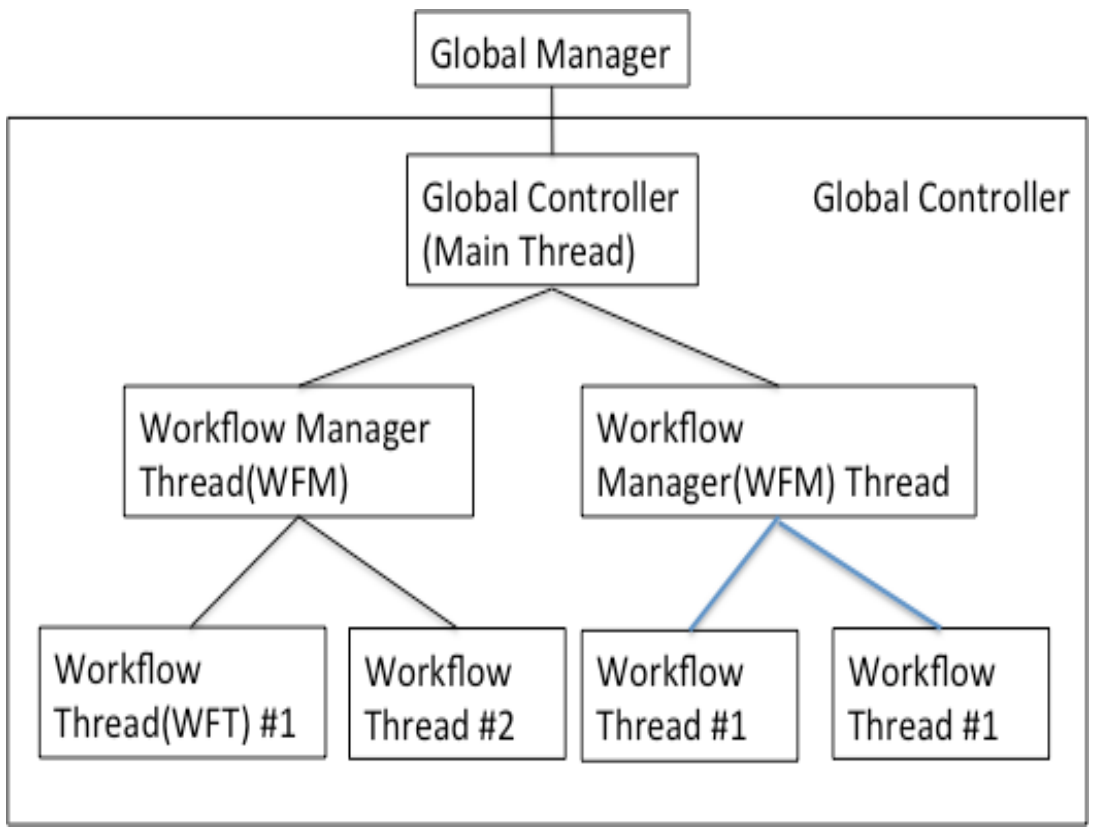

**Figure 5** The OpenADN bootstrap flow diagram

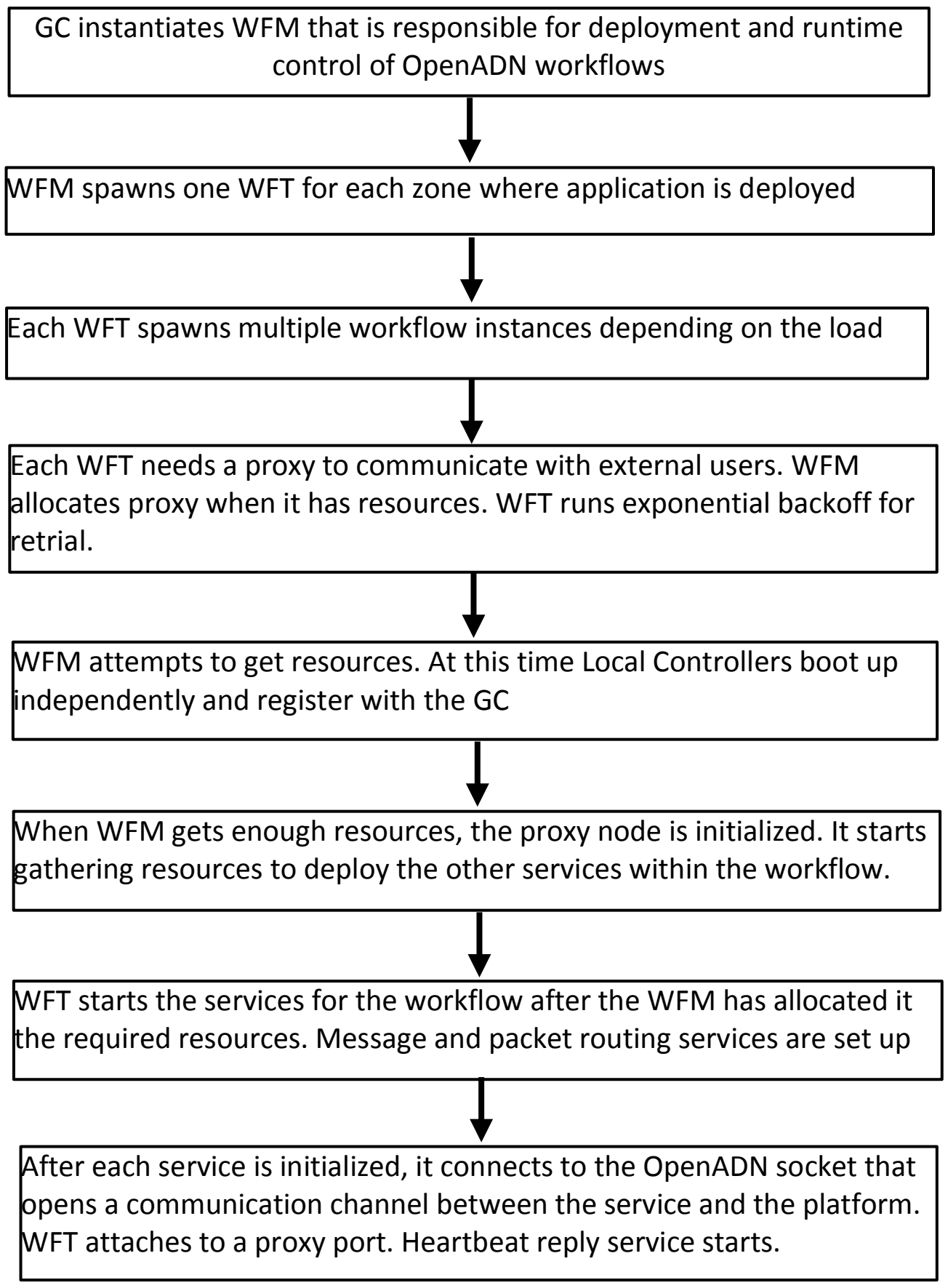

**Figure 6** Sample deterministic profile run showing creation of VMs and functions

```
Thu Apr 16 18:15:44 2015    crun1000.txt

         541337 function calls (541319 primitive calls) in 77.621 seconds

   Ordered by: cumulative time

   ncalls  tottime  percall  cumtime  percall filename:lineno(function)
        1    0.001    0.001   77.621   77.621 driver_mininet.py:2(<module>)
        1    0.849    0.849   77.613   77.613 driver_mininet.py:209(start_sim)
   111580   59.322    0.001   59.322    0.001 {built-in method poll}
        3   15.010    5.003   15.010    5.003 {time.sleep}
        1    0.001    0.001    5.183    5.183 driver_mininet.py:186(start_hosts)
        1    0.000    0.000    5.010    5.010 driver_mininet.py:163(start_fakeNameServer)
        1    0.000    0.000    5.009    5.009 driver_mininet.py:134(start_gc_lighthouseController)
     3106    0.013    0.000    1.065    0.000 driver_mininet.py:264(write)
      104    0.160    0.002    1.052    0.010 util.py:25(quietRun)
        1    0.000    0.000    0.980    0.980 driver_mininet.py:42(__init__)
        1    0.001    0.001    0.943    0.943 driver_mininet.py:67(allocate_singleSwitchTopo)
       19    0.001    0.000    0.893    0.047 node.py:300(linkTo)
       19    0.001    0.000    0.698    0.037 util.py:79(makeIntfPair)
     3106    0.671    0.000    0.671    0.000 {method 'write' of 'file' objects}
     3106    0.381    0.000    0.381    0.000 {method 'flush' of 'file' objects}
      195    0.003    0.000    0.294    0.002 node.py:235(cmd)
    40024    0.287    0.000    0.287    0.000 {time.time}
      125    0.004    0.000    0.273    0.002 subprocess.py:619(__init__)
        1    0.000    0.000    0.266    0.266 driver_mininet.py:128(start_topo)
```

**Figure 7** Sample run showing creation of local controller and client_host

```
         19    0.000    0.000    0.033    0.002 net.py:153(addHost)
         19    0.000    0.000    0.032    0.002 node.py:348(setIP)
          1    0.000    0.000    0.031    0.031 node.py:585(start)
          1    0.000    0.000    0.031    0.031 node.py:578(setup)
76711/76705    0.029    0.000    0.029    0.000 {len}
          2    0.000    0.000    0.022    0.011 moduledeps.py:60(pathCheck)
        125    0.020    0.000    0.020    0.000 {posix.fork}
          1    0.000    0.000    0.016    0.016 moduledeps.py:25(moduleDeps)
          2    0.000    0.000    0.016    0.008 moduledeps.py:7(lsmod)
          1    0.000    0.000    0.012    0.012 driver_mininet.py:175(start_client_host)
          1    0.000    0.000    0.008    0.008 driver_mininet.py:143(start_lc_lighthouse
        172    0.000    0.000    0.008    0.000 util.py:53(isShellBuiltin)
          1    0.000    0.000    0.008    0.008 node.py:634(start)
          3    0.000    0.000    0.007    0.002 node.py:450(sendCmd)
          1    0.000    0.000    0.006    0.006 net.py:560(init)
          1    0.001    0.001    0.006    0.006 util.py:1(<module>)
          1    0.000    0.000    0.006    0.006 net.py:184(addController)
```

**Figure 8** Creation of topology

```
Timer unit: 1e-06 s

Total time: 0.62231 s
File: driver_mininet.py
Function: allocate_singleSwitchTopo at line 69

Line #      Hits         Time  Per Hit   % Time  Line Contents
==============================================================
    69                                           @profile
    70                                           def allocate_singleSwitchTopo(self):
    71         1           13     13.0      0.0          yappi.start(builtins=False, profile_threads=True)
    72                                                   #pr=cProfile.Profile() #remove
    73                                                   #pr.enable() #remove
    74         1         2472   2472.0      0.4          self.switch = self.net.addSwitch('s1')
    75         1         6223   6223.0      1.0          self.net.addController( 'c0' )
    76
    77                                                   # Fake name server
    78         1         5003   5003.0      0.8          self.gc_fakeNameServer = self.net.addHost("FakeNS", mac=self.ns_mac,
    79         1        56369  56369.0      9.1          self.gc_fakeNameServer.linkTo(self.switch)
    80         1          408    408.0      0.1          print ("adding global nameserver<%s, %s, %s>"%(self.gc_fakeNameServer
```

**Figure 9** Creation of global controller

```
78    1                                 # global controller
79    1        47       47.0    0.0     mac= util.macColonHex(1)
80    1        12       12.0    0.0     ip=util.ipStr(util.ipNum(10,10,0,1))
81    1    214515   214515.0    0.3     self.gc_lighthouseController = self.net.addHost("gc" , mac=mac, ip=ip)
82    1   3405763  3405763.0    4.2     self.gc_lighthouseController.linkTo(self.switch)
83    1     36048    36048.0    0.0     print ("adding flobal lighthouse controllr <%s, %s, %s>"%(self.gc_lighthouseController.name, mac, ip))
```

**Figure 10** Profile run for linking hosts to a switch

```
121
122    6       21      3.5    0.0                   hostName = "h" + str(j-1)+ "s"+ str(i)
123    6     2362    393.7    0.4                   print ("adding host to site %s: <%s, %s, %s>" %(i, host
124    6     8833   1472.2    1.4                   host = self.net.addHost(hostName,mac=mac, ip=ip)
125    6   321058  53509.7   51.6                   host.linkTo( self.switch )
126                                                 # store info for each host
127    6       25      4.2    0.0                   hostInfo = {}
128    6        8      1.3    0.0                   hostInfo["host"] = host
129    6       24      4.0    0.0                   hostInfo["lc_controller_addr"]= lc_lighthouseControlle
130    6       11      1.8    0.0                   hostInfo["host_addr"] = host.defaultIP
131
132    6       35      5.8    0.0                   siteDesc[i-1]["hostList"].append(hostInfo)
133                                             #   self.hostList[i].append (hostInfo)
134    2        8      4.0    0.0               self.siteDescList.append(siteDesc[i-1])
```

**Figure 11** Time spent in sleep function in global controller module

```
Total time: 5.08587 s
File: driver_mininet.py
Function: start_gc_lighthouseController at line 149

Line #      Hits         Time  Per Hit   % Time  Line Contents
==============================================================
   149                                           @profile
   150                                           def start_gc_lighthouseController(self):
   151         1           16     16.0      0.0          print ("2. Starting Global Lighthouse Controller"),
   152         1          258    258.0      0.0          self.gc_lighthouseController.cmd('export HOST_NAME=%s'%(self.gc_ligh
   153         1          182    182.0      0.0          self.gc_lighthouseController.cmd('export NAME_SERVER_ADDR=%s'%(self.
   154         1          179    179.0      0.0          self.gc_lighthouseController.cmd('export NAME_SERVER_UPDATE_PORT=%s'
   155         1         6597   6597.0      0.1          self.gc_lighthouseController.cmd("python3 %s &"%(GLOBAL_CONTROLLER))
   156         1          579    579.0      0.0          print (".... started\n")
   157
   158
   159         1      5002929 5002929.0     98.4          sleep(5)
   160         1        74569  74569.0      1.5          yappi.get_func_stats().print_all()
   161         1          557    557.0      0.0          yappi.get_thread_stats().print_all()
```

**Figure 12** Time taken by polling function

```
271                                                     #start the processes in the hosts
272          1         854     854.0      0.0           print ("-------------\n")
273          1     5050107 5050107.0      2.3           simNetwork.start_fakeNameServer()
274          1         129     129.0      0.0           print ("checkpoint 1...after fakenameserver")
275          1     5085906 5085906.0      2.3           simNetwork.start_gc_lighthouseController()
276          1         126     126.0      0.0           print ("checkpoint 2 ...after gc")
277          1      330151  330151.0      0.1           simNetwork.start_lc_lighthouseControllers()
278          1        2657    2657.0      0.0           print ("checkpoint 3 ..after lc")
279          1         229     229.0      0.0           print ("\n")
280          1         213     213.0      0.0           print ("4. Starting hosts:")
281          1         286     286.0      0.0           print ("checkpoint 4...after hosts")
282          1     5465310 5465310.0      2.5           simNetwork.start_hosts()
283
284          1      468482  468482.0      0.2           simNetwork.start_client_host()
285          1        4453    4453.0      0.0           print ("-------------\n")
286          1         268     268.0      0.0           print ("checkpoint 5...after client host")
287                                                     #start the monitoring
288          1           8       8.0      0.0           endTime = time() + _runTime
289     140057     1011019       7.2      0.5           while time()< endTime:
290     140057   200834072    1433.9     90.5                   readable = poller.poll(1)
291     148007      666242       4.5      0.3                   for fd, _mask in readable:
292       7951       26630       3.3      0.0                           node = Node.outToNode[ fd ]
```

**Figure 13** Concurrency profiling output

```
2. Starting Global Lighthouse Controller .... started


Clock type: CPU
Ordered by: totaltime, desc

name                                       ncall  tsub      ttot      tavg
...7.egg/mininet/util.py:25 quietRun       179    0.372231  0.983090  0.005492
..gg/mininet/node.py:300 Host.linkTo       35     0.001244  0.770000  0.022000
..gg/mininet/util.py:79 makeIntfPair       35     0.001611  0.541088  0.015460
..on2.7/subprocess.py:757 Popen.poll       65625  0.085931  0.333069  0.000005
..ckages/line_profiler.py:95 wrapper       3/2    0.000031  0.278079  0.092693
..et.py:141 mininetDriver.start_topo       1      0.000065  0.248423  0.248423
..g/mininet/net.py:348 Mininet.start       1      0.000077  0.248282  0.248282
..ocess.py:1256 Popen._internal_poll       65804  0.152177  0.247391  0.000004
..g/mininet/net.py:303 Mininet.build       1      0.000024  0.228708  0.228708
..net/net.py:255 Mininet.configHosts       1      0.001468  0.228668  0.228668
..g/mininet/node.py:267 Host.addIntf       70     0.000413  0.226883  0.003241
..7.egg/mininet/util.py:120 moveIntf       35     0.000189  0.226470  0.006471
..py2.7.egg/mininet/util.py:91 retry       35     0.000211  0.226280  0.006465
..ininet/util.py:105 moveIntfNoRetry       35     0.001316  0.226069  0.006459
..7/subprocess.py:619 Popen.__init__       216    0.008887  0.090917  0.000421
..ocess.py:1099 Popen._execute_child       216    0.031574  0.074085  0.000343
..7.egg/mininet/node.py:235 Host.cmd       153    0.003151  0.029812  0.000195
```

**Table 1** Run time used for user and system activities

| | | | | | | | | (time unit: seconds) |
|---|---|---|---|---|---|---|---|---|
| **Run** | 1 | 2 | 3 | 4 | 5 | 6 | Average | % Run time |
| **User Space** | 0.62 | 0.59 | 0.55 | 0.63 | 0.72 | 0.74 | 0.64 | 1.45% |
| **System Calls** | 0.65 | 0.73 | 0.75 | 0.96 | 1.58 | 1.65 | 1.05 | 2.38% |
| **Run time** | 34.65 | 35.22 | 35.6 | 42.67 | 56.01 | 60.74 | 44.15 | 100 |

**Table 2** Time used for user and system activities on physical machines

| Function | User Space | System Calls | Run Time | User(%) |
|---|---|---|---|---|
| **Name Server** | 14.161 | 5.072 | 229.438 | 6.17 |
| **Global Controller** | 83.637 | 15.797 | 200.835 | 41.64 |
| **Local Controller** | 18.549 | 7.16 | 175.57 | 10.57 |
| **Node Controller** | 19.95 | 8.86 | 156.99 | 12.71 |
| **Client** | 0.428 | 0.036 | 18.855 | 2.27 |
| Total | 136.725 | 36.925 | 781.688 | 17.49 |

**Table 3** CPU Time of selected functions

| Run Time | VM Creation | VM Starting | VM Creation and Starting% | Poll | Poll % |
|---|---|---|---|---|---|
| 100.969 | 15.441 | 2.162 | 17.43 | 81.948 | 81.16 |
| 77.621 | 15.255 | 2.155 | 22.43 | 59.687 | 76.90 |
| 105.849 | 15.482 | 2.22 | 16.72 | 85.439 | 80.72 |

**Table 4** Time for different functional modules

| Module | CPU time taken (seconds) |
|---|---|
| Fakenameserver | 5.010 |
| Global controller | 5.009 |
| Local controller | 0.008 |
| Hosts | 5.183 |
| Client host | 0.012 |